\documentclass[review,12pt,sort&compress]{elsarticle}
\usepackage{times}
\usepackage{amsmath,amssymb}
\usepackage{color}

\newcommand{\removed}[1]{}

\usepackage{graphicx}

\usepackage[usenames,dvipsnames]{xcolor}
\usepackage[colorlinks=true,linkcolor=Red,citecolor=Green,linktoc=page]{hyperref}
\begin{document}

\begin{frontmatter}

\title{Coulomb blockade-tuned indirect exchange in ferromagnetic
nanostructures
}

\author{V. I. Kozub\footnote{Corresponding author, Email: \texttt{ven.kozub@mail.ioffe.ru}}}
\address{A.~F.~Ioffe Physico-Technical Institute of Russian
Academy of Sciences, 194021 St. Petersburg, Russia}

\author{Y. M. Galperin}
	\address{Department of Physics, University of Oslo, 0316 Oslo, Norway and
A.~F.~Ioffe Physico-Technical Institute of Russian Academy of Sciences, 194021 St. Petersburg, Russia}

\author{V. M. Vinokur}
\address{Materials Science Division, Argonne National Laboratory, 9700 S. Cass Avenue, Lemont, Illinois 60637, USA}

\date{\today}

\begin{abstract}
We develop a theory of the reversible switching of the magnetic state of the
ferromagnet-insulator-normal metal-ferromagnet (FINF) nanostructure.
The switching is controlled by tuning the Coulomb blockade strength via the
gate voltage on the normal metal granule. The proposed mechanism allows for realizing the switching without passing a dissipative current through the structure.
\end{abstract}


\vspace{2pc}

\end{frontmatter}

\section{Introduction}
\label{sec:introduction}

The pioneering work by Slonczewskii~\cite{Slon1} who proposed
switching of the magnetic state in the 
Ferromagnet/Normal metal/Insulator/Ferromagnet
(FNIF) multilayer structure by applying
a non-equlibrium spin current, opened a novel exciting direction in the
study of magnetic heterostructures with tunable configurations.
Following the concept of~\cite{Slon1}, subsequent papers~\cite{Schwabe,Bader}
 suggested an interesting possibility of
rotating magnetization in magnetic bilayer structures containing
an insulating layer (FIF structure) by the applied bias. It was shown that in a tunneling
bilayer, the Ruderman-Kittel-Kasuya-Yosida (RKKY) interaction
between the magnetic moments in ferromagnetic layers would oscillate
as a function of the bias. This allows for tuning the sign of the
interaction alternating between the ferromagnetic and
antiferromagnetic coupling by tuning the bias. Devices with the
bias-controlled switching magnetization are advantageous as compared
to those controlled by the external magnetic field, since the
latter is hard to localize within the small (several
nanometers size) area.  On the other hand, the bias results in
the current through a device leading to dissipation and, thus, to
power losses.

\begin{figure}[t]
\centering
\includegraphics[width=6cm]{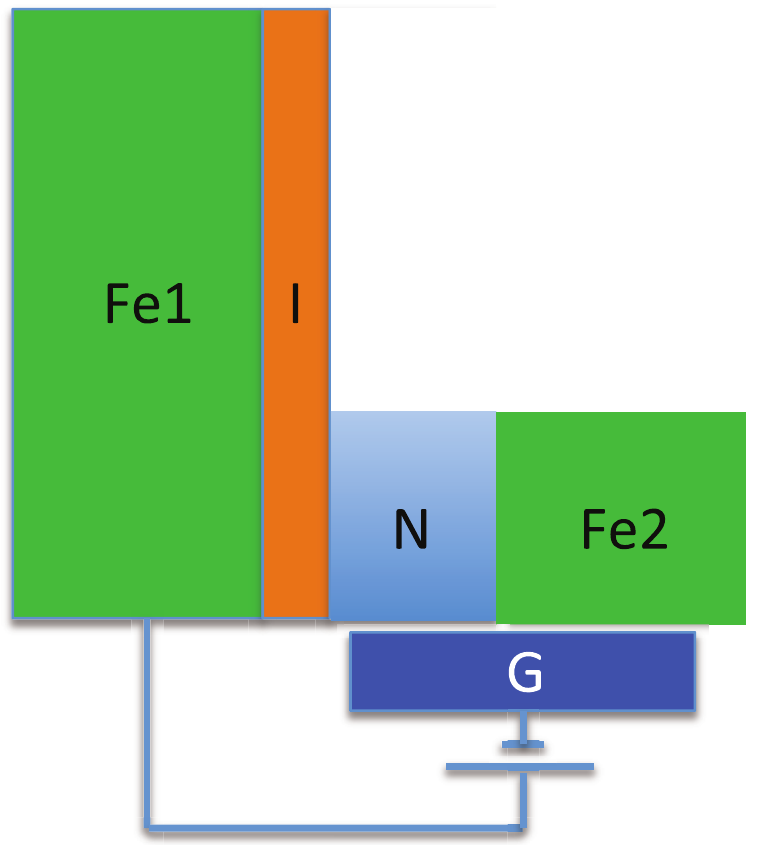}
\caption{ (Color online) 
A sketch of the experimental set up. Fe1 -- Ferromagnetic polarizer; I -- tunneling barrier; N -- normal metal layer; Fe2 -- ferromagnetic particle; G -- gate.
} \label{Fig1}
\end{figure}

In this work, building on the ideas of
Refs.~\cite{Schwabe} and~\cite{Bader} we investigate the effect of the
nanoscale tunnelling through the hybrid F1-I-N-F2 structure in which one of the ferromagnetic islands, F2, see Fig.~\ref{Fig1}, is small enough for the 
Coulomb blockade effects become essential. 
We show that in such a
structure the change of the sign of RKKY coupling 
can be achieved by tuning the electron polarization by the bias
without any current passing across the system.

\section{The Model}
\label{sec:Model}

Let us consider a multilayered system with the structure
Fe1-I-N-Fe2 (Fig.~\ref{Fig1})  where Fe1 is a bulk ferromagnet, I and
N mark insulating  and normal layers, respectively, and Fe2 is a
ferromagnetic layer shaped as
a small metal granule. In what follows we
will consider an indirect exchange between the Fe1 and Fe2
following the simple model of~\cite{Balt}. Importantly, the
indirect exchange implies the coupling between 
magnetic ions localized in different ferromagnets. The coupling is thus
 mediated by the conducting electrons whose paths traverse between
the two ferromagnets. Summation over different magnetic ions
leads to the exchange interaction between all the spins of the respective
ferromagnets, i.e., between their respective magnetizations. Hence the
interaction between the layers appears as a sum of the pair
interactions between the individual spins belonging in different
ferromagnets. In turn, the interactions between the
individual spins are described by the second order perturbation
theory with respect to spin coupling to the propagating electron modes. 
The interface between the ferromagnet and the normal metal is assumed to
be perfect, and the metal is supposed to have a simple band
structure. This model captures essential features of the
non-dissipative magnetization switch, and without any loss of
generality, we can omit the details of the interlayer coupling such
as surface imperfections, quantum well states, see,
e.g.,~\cite{cite1,cite2,cite3}.

The Hamiltonian describing the propagating electron modes $\alpha$ reads
\begin{eqnarray}
H&=&\sum_{\alpha} \varepsilon_{\alpha, b}
a^+_{\alpha,b}a_{\alpha,b} + \sum_{\alpha}
\varepsilon_{\alpha, g} a^{+}_{\alpha, g}a_{\alpha,g} 
+
\sum_{\alpha} \left(A_T a^+_{\alpha,b}a_{\alpha,g} + \textrm{h.~c.}\right)
\nonumber\\&&
+ J \sum_{j,\alpha} {\bf
S}_j{\bf s}_{e,\alpha} a^+_{\alpha,b}a_{\alpha,b} +J
\sum_{i,\alpha}{\bf S}_i {\bf s}_{e,\alpha} a^+_{\alpha, g}a_{\alpha,
g}\,.
\label{Hamiltonian}
\end{eqnarray}
Here $a^+_{\alpha, b}, a^+_{\alpha, g}$ are creation
operators for the mode $\alpha$ in the bulk and in the granule,
respectively and $a_{\alpha, b}$ and $a_{\alpha, b} $ are the corresponding annihilation
operators, $A_T$ is the tunneling amplitude, ${\bf s}_e$ is the
electron spin, ${\bf S}_j, {\bf S}_i$ are localized spins within the bulk
and in the granule, respectively, and $J$ is the exchange
integral.

Before moving further, it is instructive to compare our scheme with the previous
approaches dealing with the electron transport in nano-aggregates
including magnetic elements. The paper~\cite{martinek} reports theoretical studies
of Kondo effect in quantum dots between the two ferromagnetic
layers. The discussed phenomena are related to the
correlations between the electrons within the
ferromagnetic leads and within (non-ferromagnetic) granule, which
finally causes  an influence of these leads on the spin state of the
granule. At variance, we describe the electron-mediated
coupling between the {\it ferromagnetic ions} located within the granule
and in a  single ferromagnetic layer. This coupling has a form of 
in the effective exchange interaction
between the total spins of the layer and that of the granule,
i.e.,  between their magnetizations.  
Note, further, that while the approach by~\cite{martinek} accounts only for the
electron-electron interactions, the effect we report here 
is based on a subtle interplay between the charging energy of the granule
and the energy dependence of the electronic states mediating the indirect
exchange between the ferromagnetic layer and the granule. 
The energy dependence  of the electronic states, in turn,
transforms into the gate voltage dependence. 
As a result, the indirect exchange 
resulting from the intereference of different electron
trajectories coupling the two magnetic ions becomes dependent on the gate
voltage while the Coulomb blockade effect impedes the
current through the structure even in the presence of the gate
voltage. Note further that the design of the device of~\cite{hauptmann} is 
critically different from ours shown  in Fig.~\ref{Fig1}. The latter consists of the ferromagnetic
layer and \textit{ferromagnetic granule} separated by the passive
interlayer including the insulating and the normal-metal parts respectively, the
latter being in perfect contact with the granule. 

\section{Calculations and results}
\label{sec:Results and discussion}
\begin{figure}[t]
\centering
\includegraphics[width=6cm]{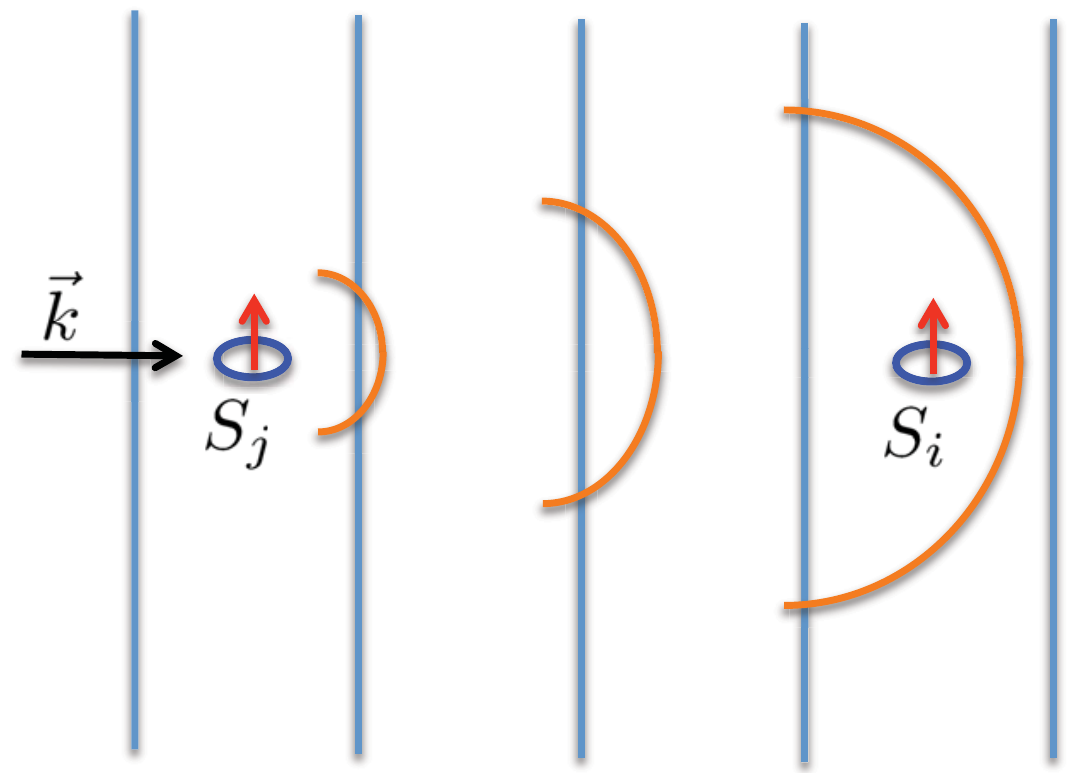}
\caption{A sketch of a general indirect exchange between the localized spins $j$ and $i$ mediated by the delocalized electrons. An incident plain wave $\bf k$ is scattered by the spin-dependent potential of the ion $j$. The interference of the incoming wave with the scattered wave at the location of the spin $i$ produces the local fluctuation of the spin density which is coupled with spin $i$.
} \label{Fig2}
\end{figure}
Starting with the Hamiltonian given by Eq.~(\ref{Hamiltonian}),
we calculate the interaction energy for the pair of
localized spins $i,j$ considering it as the result of coupling
spin $i$ to the Friedel oscillations of the spin density of
delocalized electrons produced by spin $j$~\cite{my}, see Fig~\ref{Fig2}. 
Consequently, the Friedel oscillations arise
from the interference between the unperturbed electron mode and the
scattering wave produced by spin $j$. 

\begin{figure}[h!]
\centering
\includegraphics[width=8cm]{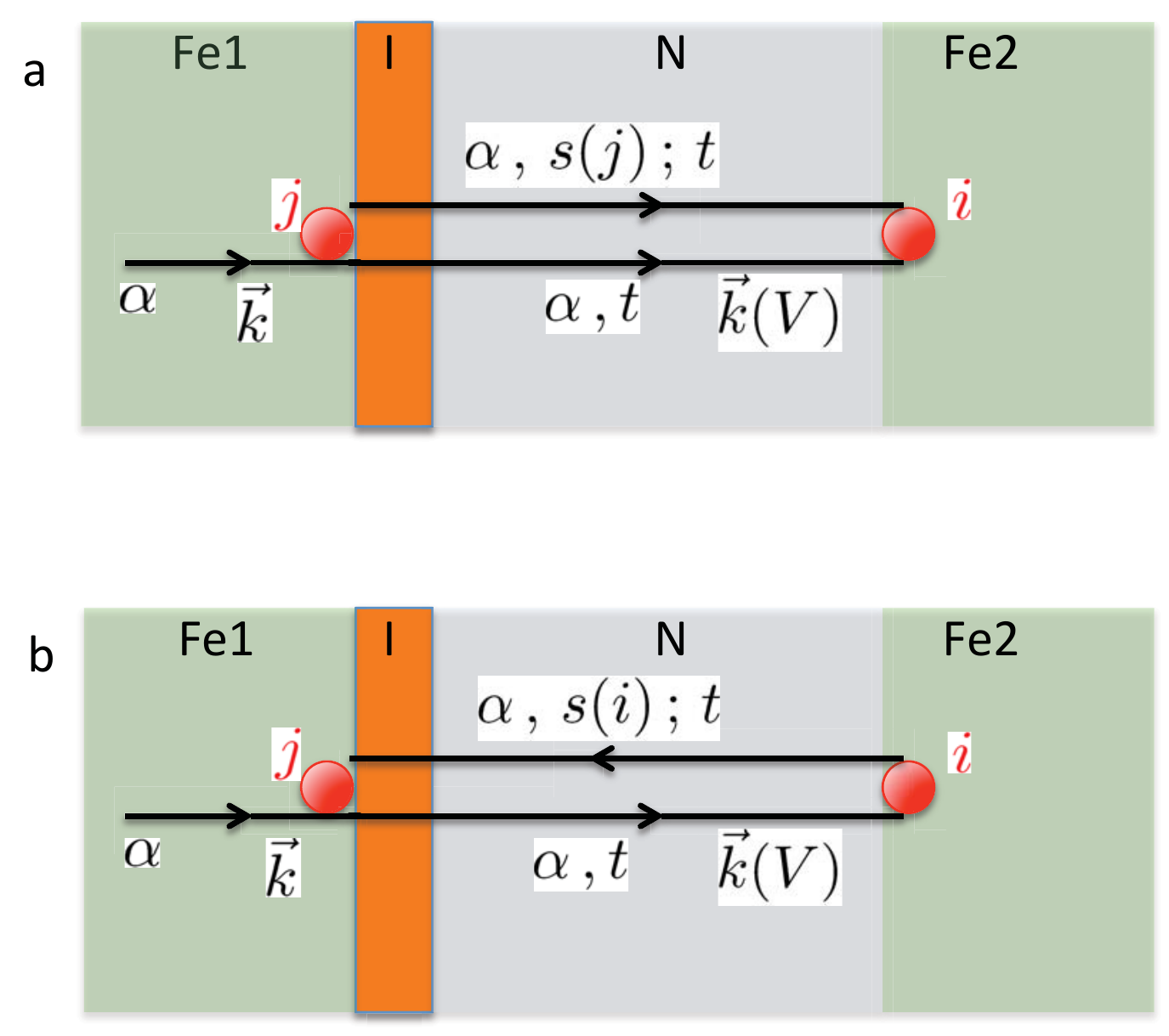}
\caption{A sketch of the processes leading to the indirect exchange between the  ions $j$ and $i$
 located at the interfaces of ferromagnetic layers F1 and F2, respectively, mediated by the incident 
 wave from the layer F1 mode $\alpha$ with the wave vector $\vec{k}$.
 (a)~The interference between the incident mode $\alpha$, propagating through
  the tunnel junction and the normal layer to the ferromagnetic layer $F2$,
  and the result of scattering of the mode $\alpha$ by the ion $j$ at the 
  point of location of the ion $i$. The index $t$ denotes the transmitted part of the wave
  confined within the normal layer, indices $s(i,j)$
 describe scattering of the mode by the respective ion. 
 The exchange occurs due to modulation of the spin density of the delocalized
 modes at the locations of the ions $i,j$. This modulation results from the 
 interference between the two different paths involving both tunneling- and
 scattering events, respectively. The dependence of the wave vector $\vec{k}(V)$ 
 on the gate voltage allows for controlling the phase of the
 interference, and, thus, controlling the sign of the exchange, by the gate voltage
 $V$. 
 (b)~The interference between the incident mode $\alpha$, and the mode,
 resulting from its scattering by the ion $i$,
 and then propagating in the opposite direction through the normal layer
 and the tunnel junction. The interference contribution is accounted at the position of the
 ion $j$.
  } \label{Fig3}
\end{figure}

Notice first, that the tunneling coupling hybridizes the states in the
bulk with the states within the granule. Thus, the
propagating mode $|\alpha; b\rangle$ of the bulk acquires, in
the first approximation with respect to $A_T$, an admixture $|\alpha;
g\rangle$ spreading within the granule and vice versa. The
boundary condition at the tunnel barrier is given as $|\alpha;
b\rangle = A_T |\alpha; g\rangle$.  Similar considerations
hold for the states $|\alpha; b, j\rangle$ resulting from
scattering of the modes $|\alpha,b\rangle$ by the spin located at the
site $ j$ within the bulk. Next, we take into account that
the state $|\alpha, g\rangle$ can suffer backscattering within
the granule at the site $ i$ forming the scattering state
$|\alpha; g,i\rangle $. 
In turn, in the second order with respect to the
tunneling amplitude, it creates an addition
$|\alpha; gran,  i;bulk\rangle$ within the bulk layer. 
This processes are shown in Fig.~\ref{Fig3}.
As a result, the propagating state resulting from the initial mode $\alpha$ assumes the form
\begin{eqnarray} 
|\alpha\rangle &=& |\alpha;b\rangle + |\alpha;g\rangle 
\nonumber \\ &&
+ \sum_j
(|\alpha;b, j\rangle +
|\alpha; b,j; g\rangle) +
\sum_i ( |\alpha; g, i\rangle + |\alpha; g,i; b\rangle) .
\end{eqnarray}
The similar procedure applies to the states $\psi_{\beta}$
which for vanishing tunnel transparency are localized within the
grain. Then any contribution of the mode $\alpha$ to the
interaction energy acquires the form $\langle\alpha; g| {\bf
S}_i|\alpha;b, j; g\rangle $. The resulting interaction energy
is obtained by the summation over the modes $\alpha$ and
$\beta$. As a result, in the lowest approximation in the tunnel
transparency, the indirect exchange coupling between the two
ferromagnetic ions located in the different ferromagnetic layers is
           \begin{eqnarray}\label{int}
           U_{\mathrm{ex}}(i,j) \! &\simeq& \!  Ja^3 
           \!\sum_{\alpha,\beta, {\bf  s}_e}
           [{\bf S_i}{\bf s}_e \psi_{\alpha ;g}({\bf R}_i)\psi_{\alpha; b,  j ;g}^*({\bf R}_i)
            + {\bf S_j}{\bf s}_e\psi_{\alpha;b}({\bf R}_{j})\psi^*_{\alpha ;g, i ; b}({\bf R}_j)
             \nonumber \\  
             &+& \!{\bf S}_j{\bf s}_e \psi_{\beta ;b}({\bf R}_j)\psi_{\beta; g,  i ;b}^*({\bf R}_j)
            + {\bf S}_i {\bf s}_e\psi_{\beta;g}({\bf R}_i)\psi^*_{\beta; b ,  j ;g}( {\bf R}_i) + \mathrm{c.~c.} ].
           \end{eqnarray}
Here $J$ is the local exchange integral magnitude, $a$ is the size of elementary cell, $\mathbf{s}_e$ is the electron
spin while $\mathrm{c.~c.}$ means complex conjunction. 
The first two terms
describe the interference of the electron states 
at sites $ i$ and $ j$, respectively, while
 the two last
terms have the similar nature, but stem from
modes $\beta$ localized within the granule. Spin dependences of the scattering amplitudes $\psi_{\alpha;b,j;g}$ are shown later, Eq.~(\ref{scam}).

One notes that the admixtures to the states $\alpha$ resulting from the
hybridization within the granule are formed from the states
$\beta$ having the same energy as modes $\alpha$. At the same
time, the tunneling of an electron from
the bulk to the granule increases the Coulomb energy of the
granule. If the gate creates a voltage $V$ between the granule and
the bulk, an electron having the kinetic energy $\varepsilon $
in the bulk, acquires the kinetic energy 
\begin{equation} \label{hyb1}
\varepsilon' =
\varepsilon - eV - e^2/C
\end{equation}
upon tunneling into the granule, where
$C$ is the granule capacitance. When a hole tunnels, the state with
the energy $\varepsilon$ of the grain is coupled to the state with
the energy 
\begin{equation} \label{hyb2}
\varepsilon'=\varepsilon - eV + e^2/C
\end{equation}
 in the bulk, see Fig.~\ref{Fig3}.
One has to bear in mind that the tunneling of a hole from the bulk to the
granule corresponds to the tunneling of an electron from the granule
to the bulk and vice versa. One also notes that the terms in Eq.~(\ref{int}) 
involving modes $\alpha$ describe tunneling of an
electron {\it to} the granule while the modes $\beta$ describe
tunneling of an electron {\it from} the granule.
One has to note further that for $|eV| < e^2/C$, the real tunneling processes are
suppressed by the Coulomb blockade and the effects become virtual.
Note, however, that the relations between the energies of the
states hybridized by tunneling given above hold irrespective to
the fact whether tunneling is real or virtual since these
relations do not depend on the occupation numbers of the relevant
states. It is the situation of virtual tunneling allowing 
eliminating direct current through the structure that will be
considered in what follows.

We chose the size of the granule to be much larger than the
electron wavelength, this implies that the states with a given
wave vector can be considered as the basis. 
Note that while
one could have been expecting
that $\alpha$-modes  correspond to $k_x > 0$ and
the $\beta$-modes correspond to $k_x < 0$, this appears not be the case.
Indeed, let us consider an electron mode with $k_x < 0$ incident from
the granule to the interface with the tunneling layer. The mode is partly
reflected back to the granule acquiring $k_x > 0$ and partly tunnels to
the bank. The following interference schemes can be realized: (i)
The wave scattered  from the ion $j$ in the bulk tunnels back to
the granule and interferes on the ion $i$ with the wave reflected
from the interface, (ii) the reflected wave is scattered by the ion
$i$ and the scattered wave tunnels to the bank where it interferes
on the ion $j$ with the tunneling tail of the initial incident
mode. One notes that the corresponding interference terms coincide
with the ones in Eq.\,(\ref{int})  which result from the modes
$\alpha$ with $k_x > 0$ incident from the bulk. The same
considerations can be applied to the modes $\alpha$ which acquire
the terms with $k_x < 0$ due to acts of reflection from the back
boundary of the granule. These additional terms in Eq.\,(\ref{int})
are included as complex conjunction of the first four terms.

The propagating exponentials entering $\psi_{\alpha}, \psi_{\beta}
$ have a form $\psi = a^{-3/2} \exp ( i {\bf k R})$ where we have
used a normalization with respect to the volume of an elementary
cell $a^3$. The scattering amplitudes, $\psi_{\alpha;b,j;g}$, in the Born
approximation are given by
\begin{equation} \label{scam}
\psi_{\alpha;b,j;g}({\bf R}_i,{\bf R}_j ) = \frac{A_T J {\bf S}_j{\bf s}_e
ma^3}{2\pi\hbar^2|{\bf R}_j - {\bf R}_i|a^{3/2}}\exp \left[ i({\bf k}{\bf
R}_j + k|{\bf R}_j - {\bf R}_i|)\right ] \,.
\end{equation}
For other scattering amplitudes one has similar expressions
differing by the notations of the scattering ions and by the signs
of the wave vectors. Finally inserting the corresponding
amplitudes to Eq.~(\ref{int})  after the
summation over the electron spins one obtains for the exchange coupling the expression  $U_{\mathrm{ex}} ({\tilde {\bf R}_i} - {\tilde {\bf R}_j})$ where
  \begin{eqnarray}\label{int1} 
       {U_{\mathrm{ex}}}({\tilde {\bf R}})  &=& J^2{\bf S}_i{\bf S}_j\frac{ma^3| A_T |^2}{2\pi \hbar^2 \tilde
       { R}}
       \nonumber \\ && \times
              \sum_{{\bf k}, \nu} 
              \left(
        2 e^{i ( {\bf k'}_{\nu} {\tilde {\bf R}}  -  {k'}_{\nu}{\tilde { R}})} 
         +    e^{i (- {\bf k'}_{\nu} {\tilde {\bf R}}  -  {k'}_{\nu}{\tilde { R}})}
         +    e^{i (- {\bf k'}_{\nu} {\tilde {\bf R}}  +  {k'}_{\nu}{\tilde { R}})} 
             \right)
       \end{eqnarray}       
 where $\mathbf{k}'$ is the wave vector for the state with the kinetic energy
$\varepsilon'$ (thus depending on the gate voltage). Here we
redefined the spatial coordinates, $\tilde {\bf R}$, extracting
the thickness of the insulating layer, $\tilde{ R} \equiv |\tilde {\mathbf{R}}|$.
Indices $\nu = 1,2$ stand for the
electron and hole channels, respectively.  In general, the factor
$|A_T |^2$ depends on the electron 
wave vector  $\bf k$. However
this dependence is relatively weak as compared to that resulting
from the strongly oscillating exponential and does not influence
the effect of the sign change. Thus, in the lowest approximation one
can neglect the dependence of $A_T$ on the  $\bf k$ and the
applied voltage. 

Let us first perform an integration over directions of $\mathbf{ k}_\nu '$ which enter through the 
combinations $\exp(ik'_\nu \tilde R \cos \theta)$. 
where $\theta$ is an angle between the 
vectors $\mathbf{k}'_\nu$ and $\tilde {\mathbf{R}}$. The
integration over $\theta$ of the exponential given above yields 
$(2/k'_\nu)\sin (k'_\nu \tilde R)$
Performing similar integration of all the terms entering r.h.s. of Eq.~(\ref{int1})
we obtain
\begin{equation}
{U_{\mathrm{ex}}}(\tilde {R}_{ij})  = J^2{\bf S}_i{\bf S}_j\frac{ma^3| A_T
|^2}{\pi \hbar^2 \tilde
      { R}_{ij}^2}
              \sum_{ k', \nu}\frac{\sin 2k'_\nu \tilde {R}_{ij} }{k'_\nu}.
\end{equation}
Now let us perform an integration over $k$. Based on the
relations (\ref{hyb1}) and  (\ref{hyb2}) between the energies of electronic states hybridized by
tunneling we have:
        \begin{equation}\label{exch2}
                 { k'}_1 = k\left[ 1 - \frac{1}{2}\left(\frac{eV + e^2/C}{\varepsilon}\right)\right] ,\quad
                   { k'}_2 = k\left[ 1 -
               \frac{1}{2}\left(\frac{eV - e^2/C}{\varepsilon}\right)\right]\,.
              \end{equation}
It is important that an electron can tunnel only {\it from} an
occupied state while the hole can tunnel only {\it to} an occupied
state. Accordingly, for $\nu = 1$ the integration over $k$ is up
to $k_F$, while for $\nu = 2$ integration goes up to 
\begin{equation}\label{kh}
k_h =
k_F\left (1 + \frac{eV - e^2/C}{2 \varepsilon}\right).
\end{equation}
Correspondingly, we have
\begin{eqnarray}\label{fin} 
{U_{\mathrm{ex}}}(\tilde R_ {ij})  &=& J^2{\bf S}_i{\bf S}_j\frac{2ma^3| A_T|^2}
{2\pi \hbar^2k_F^2\tilde R_ {ij}^3}
\nonumber \\ &&  \times
\left\{\cos \left[2k_F\left(1 -  \frac{eV +
       e^2/C}{2\varepsilon}\right)\tilde R_ {ij} \right ]+ \cos ( 2k_F\tilde R_ {ij} )\right\} \! .
\end{eqnarray}
Note that the dependence on the gate voltage
exists only in the electron channel describing an electron
tunneling to the granule and is absent in the hole channel where
an electron is tunneling {\it from the granule} and thus its
momentum within the layer of a normal metal is not affected by the
gate voltage since this layer is located {\it within the granule
itself}.

 Next, making use of Eq.~(\ref{fin}) we employ the calculation procedure
similar to that of \cite{Balt}.  Namely, we assume that our
granule is a slab including intermediate normal layer and
separated from the bulk ferromagnet by a tunnel barrier (the
latter is a new feature as compared to the model of
Ref.\,\cite{Balt}). We replace the summation over ion $j$ by
integration over corresponding spatial coordinates and take
into account that the integrand strongly oscillates with the spatial
scale of the order of an elementary cell size $a$. As a result, only
the interface ions of the feromagnet contribute efficiently. Thus
only spins $i$ corresponding to the interface of the ferromagnet
are important. Thus we perform an integration over the volume
of Fe2  (granule) for the each ion $i$ at the
interface of the Fe1  with the area $A$.

Note that the most efficient coupling could be expected for the
case where the oscillations of the indirect exchange do not
cancel each other, that is for the granule of an atomic size. However,
at present, the working body of realistic devices on the base of ferromagnetic metals
are by far exceeding the atomic dimensions.

Finally, we arrive at the following expression for the coupling between
the particles per unit area:
\begin{equation}\label{E}
{U_{\mathrm{ex}} \over A} = \frac{J^2 m  S_g S_b}{16 \pi^2 \hbar^2} K(k_Fd)\,,
\end{equation}
where $d$ is a distance between the ferromagnetic layers, $A$ is
the contact area, $S_g$ and $S_b$ are the values of localized
spins within the bulk and within the grain, respectively, and
%
\begin{equation}\label{array} 
                  K (z) =B\frac{|A_T|^2 }{ z^2}
                  \left\{\sin z + \sin\left[
                  z\left (1 -  \frac{eV +
       e^2/C}{2\varepsilon_F}\right)\right]\right\}\,,
\end{equation}
with $z = 2k_F d$, $B$ being the numerical factor of the order of
unity, and $d$ being a thickness of the normal metal layer. In the course
of our derivation we exploited the fact that $k_F \simeq \pi/a$.
For small $eV$ one finds
         \begin{equation*} 
                K (z)\simeq  \frac{|A_T|^2}{z^2} \left\{\sin z  + \sin \left[ z \left ( 1
                - \frac{e^2}{2C\varepsilon_F}\right)\right] - \frac{eV}{2\varepsilon_F} \cos \left[z\left( 1
                - \frac{e^2}{2C\varepsilon_F}\right)\right]\right \}\,. 
              \end{equation*}
Thus, 
the voltage-dependent part is 
        \begin{equation}\label{finend}
                 K (z) \simeq -\frac{eV|A_T|^2}{2\varepsilon_Fz}  \cos\left[ z \left(1 -
                \frac{e^2}{2C\varepsilon_F}\right)\right]\,.
        \end{equation}
The 
 sign of this contribution to the  exchange energy is changed with the sign of the voltage.
Note that the effect we describe is completely different from the
mechanisms of switching suggested before in
\cite{Schwabe,Bader,Slon1} since \textit{it implies no current
through the device and therefore, no dissipation}.

\section{Discussion and estimates} \label{discussion}
One of the most promising ways for an experimental realization of the
proposed structure is the point contact nanofabrication technique (see,
e.g.,~\cite{Ralph,Ralph2,Ralph3}). It allows, in particular,
manipulating with ferromagnetic or normal metal granules with a size
of a few nanometers  \cite{Ralph,Ralph2}, or even with single
molecules \cite{Ralph3}. Manipulation with metallic layers with a
thickness $\sim 1$~nm within the point contact was reported in
\cite{Ralph2} while a presence of tunnel barriers \cite{Ralph2,Ralph3}
can be used to fabricate gated structures. We suggest
that the experimental system is to be produced on the base of a ferromagnetic
granule covered by a normal metal layer fabricated within the
nanoscale point contact between ferromagnetic (Fe1) and normal
electrodes, the latter playing a role of the gate. The magnetic
state of the granule can be probed by application of short pulses of
large bias (lifting the Coulomb blockade)
 since the current through the structure is
sensitive to mutual orientation of magnetization in Fe1 and Fe2.

For a typical size of the grain of 5 nm one estimates $e^2/C \sim
0.1$ eV and $k_F d \sim 10 $. Thus the Coulomb blockade regime
exists until $V < 0.1$~V while the coupling in this regime
according to Eq.~(\ref{array}) oscillates with the variation of
$V$. 

Let us evaluate the relative efficiency of the
bias-controlled exchange with respect to the magnetic dipole
interaction. In typical ferromagnets the ratio of the dipole
interaction, $E_{d} \sim \mu^2/a^3$, where $\mu$ is a magnetic moment
of a magnetic ion while $a$ is a lattice constant,  to direct
exchange energy, $E_{\mathrm{ex}}$, for neighboring spins is of the order of
$10^{-4}$. The effective magnetic field near the surface of the
larger ferromagnet is $H_D \sim \gamma (\mu/a^3)$ where $\gamma$
is demagnetizing factor. For a slab with a thickness $b$ and a
linear size $c >> b$ it is of the order of $b/c << 1$. Thus the
total dipole energy is of the order of $H_D \mu \mathcal{V}/a^3$ where $\mathcal V$
is a volume of the smaller ferromagnetic particle. If the latter
is a slab with a thickness $t$ and a linear size $L$, $\mathcal{V} \sim
tL^2$. Correspondingly, the dipole energy $E_D \sim \gamma E_d
(tL^2/a^3)$.

Making use of Eqs.~(\ref{fin}) and (\ref{array}), one estimates the
efficiency of the bias-controlled exchange for
$(eV/\varepsilon_F)k_Fd > 1$ as $E_X \sim
U_{\mathrm{ex}}(L/a)^2 |A_T|^2 /(k_Fd)^2$. Correspondingly,
\begin{equation}
\frac{E_X}{E_D} \sim \gamma^{-1} \left(\frac{U_{\mathrm{ex}}}{E_d}\right)
\left(\frac{a^5}{d^2tL^2}\right)\left(\frac{L^2|A_T|^2}{a^2}\right)\,.
\end{equation}
The Coulomb blockade conditions imply that $|A_T|^2(L/a)^2 < 1$.
As a result, the bias-controlled exchange dominates the dipole
interactions provided
\begin{equation}
\gamma^{-1}\left(\frac{U_{ex}}{E_d}\right)
\left(\frac{a^5}{d^2tL^2}\right) > 1\,.
\end{equation}
Taking $t \sim L$, $d \sim  1$~nm one sees that this condition is
satisfied if $L \leq 5$~nm provided $\gamma = 10^{-1}$ and is
experimentally realistic \cite{Ralph,Ralph2,Ralph3}.

The switching of the granule magnetization can be, in
particular, registered due to Giant Magnetoresistance effect
studying a (weak) direct current through the structure. As it is
known, the resistance of the spin valve of the sort of studied in
our paper is sensitive to mutual orientation of magnetizations of
the magnetic layers.

The following note is in order. Our simplified approach does not
take into account the surface imperfectness and realistic band
structure, and we present only the order of magnitude estimates.
Yet, our conclusions that are based on the facts that (i) the RKKI
interaction strongly oscillates as a function of distance between
the interacting spins and (ii) the period of the oscillations
depends on the actual electron wave vector which can be affected by
the applied electric field are not affected by the
details the surface and band structures.

\section{Summary}
\label{sec:Summary}

We have proposed a theory of the effect
of the dissipationless magnetic coupling in a hybrid structure
consisting of two ferromagnetic layers separated by a tunnel
barrier and normal metal, which utilizes the Coulomb blockade
effects, suppressing the real tunneling processes, and
realizes the switch controlling the sign of the coupling.

\section*{Acknowledgments} 
This work is supported by the U.S. Department of Energy, Office of Science, Materials Sciences and Engineering Division.

\section*{References} 


\end{document}